\documentclass[conference]{IEEEtran}
\IEEEoverridecommandlockouts
\usepackage{cite}
\usepackage{amsmath,amssymb,amsfonts}
\usepackage{algorithmic}
\usepackage{graphicx}
\usepackage{textcomp}
\usepackage{verbatim}
\usepackage{color}

\def\BibTeX{{\rm B\kern-.05em{\sc i\kern-.025em b}\kern-.08em
    T\kern-.1667em\lower.7ex\hbox{E}\kern-.125emX}}
\begin{document}

\title{Managed Blockchain Based Cryptocurrencies with Consensus Enforced Rules and Transparency}

\author{
\IEEEauthorblockN{Peter Mell}
\IEEEauthorblockA{National Institute of Standards and Technology\\
Gaithersburg, Maryland 20899\\
Email: peter.mell@nist.gov}
}

\maketitle

\begin{abstract}
Blockchain based cryptocurrencies are usually unmanaged, distributed, consensus-based systems in which no single entity has control. Managed cryptocurrencies can be implemented using private blockchains but are fundamentally different as the owners have complete control to do arbitrary activity without transparency (since they control the mining). In this work we explore a hybrid approach where a managed cryptocurrency is maintained through distributed consensus based methods. The currency administrator can perform ongoing management functions while the consensus methods enforce the rules of the cryptocurrency and provide transparency for all management actions. This enables the introduction of money management features common in fiat currencies but where the managing entity cannot perform arbitrary actions and transparency is enforced. We thus eliminate the need for users to trust the currency administrator but also to enable the administrator to manage the cryptocurrency. We demonstrate how to implement our approach through modest modifications to the implicit Bitcoin specification, however, our approach can be applied to most any blockchain based cryptocurrency using a variety of consensus methods.
\end{abstract}


\begin{comment}
LEE BADGER COMMENT:
It would be great to have some kind of state machine diagram, with preconditions on state transitions, that conveys how the modes work. As is, the details are in English and inside paragraphs. I think I followed it, but I have less confidence as a reader than I would if there were a state machine formulation.
\end{comment}

\section{Introduction}
\label{Introduction}
Blockchain based cryptocurrencies are usually unmanaged, distributed, consensus-based systems in which no single entity has control \cite{Baliga2017}. They use open consensus based approaches that allow anyone to participate in maintaining the blockchain, even retaining their anonymity. Such systems remove the need for a third party in financial transactions and eliminate the double spending problem (where the same digital cash is spent multiple times) \cite{Swan2015}. This lack of a need for a trusted third party is supposed to result in reduced transaction fees over non-cryptocurrency based systems (e.g., credit cards), enabling efficient micropayments \cite{Narayanan2016}. Recently however, limitations with some cryptocurrencies on transaction throughput has caused transaction fees to be high. Lastly, such systems generally provide a level of anonymity where individuals are not linked to accounts and where it is trivial for an individual to produce and use new accounts. Examples of such systems include Bitcoin \cite{Nakamoto2008}, Ethereum \cite{wood2014ethereum}, Bitcoin Cash \cite{BitcoinCash}, Litecoin \cite{Litecoin}, Cardano \cite{Cardano}, NEM \cite{NEM}, Dash \cite{Duffield2014}\footnote{Any mention of commercial products is for information only; it does not imply recommendation or endorsement. The blockchain based cryptocurrencies listed are the ones with the largest market capitalization in descending order as of 2017-12-29 according to \cite{Coinmarketcap}.}. 

In this work, we consider how to bring many of the advantages of such open consensus based cryptocurrencies to the area of managed cryptocurrencies\footnote{Note that managed cryptocurrencies also use consensus methods but they are not open to public participation.}. We refer to a currency as `managed' if there exists an owner that can exert control over the currency. Managed currencies include electronic representations of fiat currencies as well as virtual world and in-game currencies. In the cryptocurrency realm, they are often referred to as `permissioned blockchains' (examples include Multichain \cite{Greenspan2015} and Ripple). With managed currencies, the identity of individuals is often, but not necessarily, linked to the accounts (e.g., as when someone opens a bank checking account). Furthermore, the managing entity usually reserves the right to control the money supply (i.e., they can print money). And law enforcement related functions may include freezing or confiscating assets. Managed cryptocurrencies can be implemented with private blockchains using tools such as Multichain. However, in such implementations the owners have complete control to perform arbitrary activity without transparency. This is because the owners authorize (and thus control) the servers maintaining the blockchain.

In our research we explore a hybrid approach where we merge strengths of open consensus based cryptocurrencies with features often found in managed currencies. In doing so we design not a particular cryptocurrency, but instead a flexible architecture that allows for different implementations. From the open consensus approach we leverage the ability of the mining community to enforce the rules of the currency and to enforce transparency, where all transactions are publicly viewable. In this way the managing entity of the cryptocurrency cannot perform arbitrary actions, but only those explicitly allowed in the cryptocurrency design and all such management actions are publicly recorded in the blockchain. From the managed currencies, we leverage concepts such as the ability of the currency administrator to create funds, tie user identity to accounts, freeze/confiscate funds (e.g., due to illegal activity), and set the block awards for miners. This last feature indirectly enables the currency administrator to control the electricity consumption of the consensus mechanism (since fewer miners will participate if the rewards are lower). Energy consumption has often been cited as a major problem with consensus 'proof-of-work' systems; in 2014 Bitcoin mining consumed as much electricity as Ireland\cite{2014bitcoin}.

Since our approach is an architecture, the creator of any particular managed cryptocurrency instance can choose which features to include or exclude. Our architecture is flexible such that it can be used to implement open consensus environments like Bitcoin as well as closed controlled environments achievable with systems like Multichain. However, our approach is not intended for that purpose. Our area of interest is where the architecture is used to create hybrid approaches that combine the strengths (and weaknesses) of both. Note that we are not advocating any particular approach in this work and our goal is not to propose the creation of any specific cryptocurrency. Rather, we explore here the technological foundations that can enable the merging of the managed cryptocurrency idea with an open consensus based architecture and explore the resultant strengths and weaknesses. 

To enable management of the currency, we propose using a genesis transaction. All blockchains have a genesis block which is the first block, but this genesis transaction is a first transaction from which all subsequent transactions are authorized. The genesis transaction authorizes a special root account that has the currency manager role and that will be controlled by the currency administrator (the entity issuing the cryptocurrency). Our tagging of accounts with roles is key to our architecture. Accounts with the currency manager role can configure the currency to have different properties through defining policy (e.g., adjusting the roles implemented and mining rewards). Also, these accounts can issue transactions to create other accounts with different roles, in a hierarchical fashion with accounts closer to the root being more authoritative. The possible roles include currency manager, central banker, law enforcement, user, and account manager. The central bankers can create and delete funds. Law enforcement can freeze account and confiscate funds (e.g., for fraudulently gained funds being sent to terrorist organizations \cite{Lee2017})\footnote{Note that in most consensus based cryptocurrencies, restoration of funds is impossible without forking the currency.}. Users can perform monetary transactions without the need for a trusted third party. And account managers can create user accounts (and may be required to link them to physical identities).

We demonstrate how to implement our approach through modest modifications to the implicit Bitcoin specification. We chose Bitcoin because it is was the first blockchain based cryptocurrency and is the most used. However, our approach can be applied to most any blockchain based cryptocurrency (including smart contract approaches such as Ethereum). We modify Bitcoin as little as possible to facilitate implementation of our specification; all of our features were implemented through small changes to the Bitcoin transaction format. Currency managers can issue policy in such a way that the changes are reversible or permanent. Permanent changes restrict the currency manager's future actions (since they cannot be undone). Such changes are important as they can provide users confidence in the system through knowledge that the currency administrator will abide by a set of self-established rules. Added to this, the architecture requires that all management actions be transparent to the users.

Key to this approach are our solutions for maintaining a balace of power. The consensus based methods must ensure that the currency administrator (who owns the root currency manager node) abides by the stated rules of the cryptocurrency and enforces transparency of all management actions. However, the participants in the consensus methods should not be able to take control away from the currency administrator nor exclude any management transactions from entering the blockchain.

In summary, open consensus based unmanaged cryptocurrencies provide significant new benefits over previous electronic cash efforts. They eliminate the need for trusted third parties by eliminating the double spending problem, remove the need for a dedicated and centralized infrastructure, and allow for the possibility of very low transaction fees thus enabling inexpensive micro-transactions
\footnote{Bitcoin has high transaction fees due to limits on transaction throughput, but this is a technical problem not necessarily present in other cryptocurrencies.}. 
However, this model is unsuitable for managed cryptocurrencies because it is completely controlled by whomever joins the cryptocurrency network to maintain the blockchain (an open and anonymous group). Previous efforts to support managed cryptocurrencies have used permission-based blockchains where the administrators can control all access to the blockchain, ability of users to issue transactions, and ability of miners to maintain the blockchain. This is a powerful and efficient paradigm for many use cases. However, the user base must have complete trust in the currency administrator. In our work, we are attempting to eliminate the need for users to trust the currency administrator but also to enable the administrator to manage the cryptocurrency. At the same time, we are attempting to incorporate the many benefits achieved by unmanaged cryptocurrencies while mitigating the weaknesses (especially in the area of power consumption in maintaining the blockchain).

The main deliverable this paper is a novel architecture for maintaining a managed cryptocurrency through distributed consensus based approaches (eliminating the need for users to trust the currency administrator), as well as an evaluation of the resultant benefits and weaknesses. It also provides technical bit-level details on how to modify the Bitcoin specification in order to implement the approach. In future work, we will provide such an implementation and perform empirical studies. We expect the necessary code changes to be relatively straightforward given our modest changes to the specification, but this cannot be claimed until a prototype implementation has been developed.

%The rest of this work is organized as follows. Section \ref{Related Work} discusses the related research. Section \ref{Architecture} presents the overall architecture and design objectives. Section \ref{Bitcoin Specification} provides an overview of the Bitcoin specification. Section \ref{Technical Design} discusses the technical details (our modifications and additions to the Bitcoin specification). Section \ref{Security} describes our security models and associated threat mitigations. Section \ref{Analysis} analyzes the resulting capabilities, advantages, and limitations of our architecture. Section \ref{Conclusion} concludes.

\section{Related Work}
\label{Related Work}

To our knowledge, this is the only work combining the idea of a managed cryptocurrency with the open consensus model used by unmanaged currencies. The work most similar to ours is Multichain. It provides a platform for creating and deploying `private' blockchains within or between organizations. It is designed to provide the following features \cite{Greenspan2015}:
\begin{enumerate}
\item `to ensure that the blockchain's activity is only visible to chosen participants'
\item `to introduce controls over which transactions are permitted'
\item `to enable mining to take place securely without proof of work and its associated costs'
\end{enumerate}

Instances of Multichain have an administrator or group of administrators that define the ongoing policy of the system. They have complete control in defining who can view the blockchain, who can put transactions on the blockchain, and who can maintain the blockchain (those mining new blocks). This last feature enables them to maintain the blockchain at very little cost since the computationally expensive proof-of-work consensus methods of Bitcoin can be dispensed with. This is replaced with a flexible round robin approach where the miners mostly take turns publishing the new blocks and generally do not receive any reward for doing so (since the work is trivial).

While a powerful approach for organization-run blockchains, Multichain cannot be used to satisfy our stated objectives since the administrators have complete control. There is no mechanism to implement a balance of power where the administrators can manage the currency in an ongoing fashion but where the maintainers of the blockchain can ensure that the administrators follow the stated rules of the cryptocurrency. 

Country specific managed cryptocurrencies exist or are in the process of being deployed, not all of them being blockchain based, and the degree to which they are `managed' varies greatly. Dubai has launched its own cryptocurrency called emCash \cite{Buck2017}. Singapore has announced experimentation with one \cite{Cheng2017} and Estonia has announced thier `estcoin' \cite{Korjus2017}. The company Monetas \cite{Monetas} offers a product to enable countries to issue their own digital currencies; it is being actively used by several countries. Senegal is piloting a digital currency called eCFA using the Monetas platform that, if successful, will be used by Cote d'Ivoire, Benin, Burkina Faso, Mali, Niger, Togo and Lusophone Guinea Bissau \cite{Chutel2016}. Tunisia has done the same using the Monetas platform \cite{Smart2016}. The Russian Central Bank has publicly pushed for a national cryptocurrency \cite{Helms2017}. Venezuela has announced that it will launch an oil-backed cryptocurrency \cite{AlexandraUlmer2017}. And lastly, the Bank for International Settlements released a report noting that countries may need to replace cash with national cryptocurrencies \cite{Cheng2017}.

In the area of unmanaged cryptocurrencies, there exist hundreds of them. Bitcoin was the first to use blockchains and was introduced in 2008 \cite{Nakamoto2008}. There exist many forks and variants of Bitcoin, mostly optimizing certain features but often introducing novel and revolutionary architectural changes. 
We review here the blockchain based cryptocurrencies with the largest market capitalization, as of 2017-12-29. Ethereum was the first production product to enable executable programs (called smart contracts) to be put on a cryptocurrency blockchain \cite{wood2014ethereum}. Ripple \cite{Ripple} provides a solution for banks to send payments globally. Bitcoin Cash \cite{BitcoinCash} is a fork of Bitcoin with a much larger block size limit. This enables many more transactions per block thereby increasing throughput and driving down transaction fees. Litecoin \cite{Litecoin} is almost identical to Bitcoin but with several differences: smaller block publication time, larger maximum number of coins, and a change in hashing algorithm. Cardano \cite{Cardano} is based on \cite{Kiayias2017} describing a `provably secure proof-of-stake blockchain protocol'. NEM \cite{NEM} incorporates a reputation system, proof-of-importance, and multisignature accounts. Dash \cite{Duffield2014} is `privacy-centric' with a two-tiered administration network and an ability for users to instantly send coin.

\section{Managed Cryptocurrency Architecture}
\label{Architecture}

All blockchains contain a `genesis block'. This is the first block on the blockchain and it has no pointer to a previous block (being the first one). All users of the blockchain must agree on this first block for a consistent view of the blockchain to exist. We propose the addition of a `genesis transaction'\footnote{This is related to the "asset genesis" metadata transaction idea \cite{Greenspan2015} but is more powerful as it controls all transactions on the blockchain.}. This is the first transaction in the blockchain and it defines an account that has the currency manager role (and is owned by the currency administrator). In our system, only accounts with roles can issue transactions and only accounts with the currency manager role can create other accounts with roles (with one important exception, discussed later). Thus, the genesis transaction is the transaction that enables all other transactions.  

The initial account is the root of a hierarchical tree of nodes, where each node represents an account labeled with a set of roles\footnote{We use the terms node and account interchangeably depending upon the desired perspective (node in a tree versus account owned by a user)}. The root node not only has the currency manager (M) role\footnote{The M role is distinct from the currency administrator. Many accounts may have the M role but there exists a single entity which is the currency administrator.}, but it has all other available roles: central banker (C), law enforcement (L), user (U), and account manager (A). We label the roles of an account by concatenating all applicable labels. Thus, the root node has the role set `MCLUA'. 

When a node with the M role creates a new account (more precisely, it labels some unlabeled account created by some user), it bestows on that account a, not necessarily proper, subset of its roles. Thus, the cardinality of the set labels for nodes monotonically decreases as one traverses higher in the hierarchy tree. One exception to this monotonicity rule is that nodes with the M label may also modify the role sets of nodes higher in the tree (provided they are on the path from the target node to the root), restricted again to the set of roles possessed.

Nodes with the A role may also create and delete accounts, but such created accounts may only have the U role. The currency administrator then can delegate user account management to third party organizations by giving them the A role.

The different roles provide different accounts different capabilities:

\begin{itemize}
\item 
The U role enables an account to receive and spend coins. An account for which the U role has been removed has its funds frozen. 
\item
The A role enables an node to create accounts with the U role (and only the U role). It may also remove the U label for its descendants. 
\item
The C role enables the creation of new coins (apart from the block mining rewards). 
\item
The L role enables an account to forcibly move funds between accounts, to remove the U label, and to restore a previously removed U label. However, these actions can only be performed against nodes with the same or greater distance from the root. 
\end{itemize}

The currency administrator, who will own the root M labelled node, may require that A nodes verify users' identities prior to providing an account. In this case, the architecture enables a system where the `know your customer' (KYC) laws might be satisfied. Individual transacting parties would not know each other's identities but some account authorizing entity would have a record for each account with the U role. Fulfilling KYC laws is a general problem for cryptocurrencies \cite{Staples2017}. 

\begin{figure}
\centerline{\includegraphics[scale=.2]{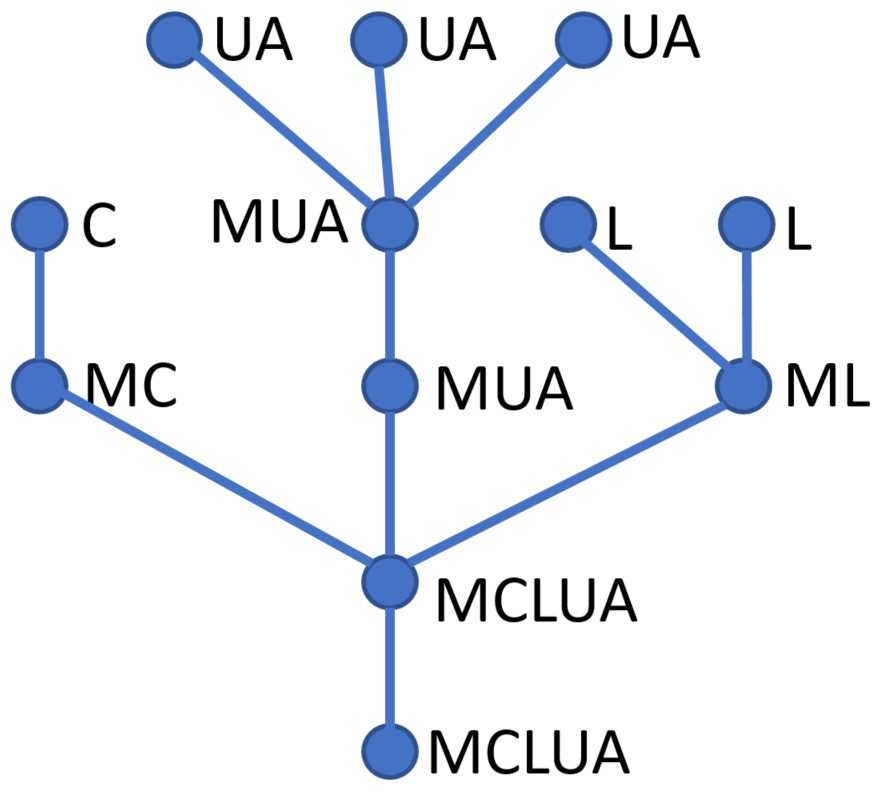}}
\caption{Example Managed Cryptocurrency Hierarchy.}
\label{fig:hierarchy}
\end{figure}

Figure \ref{fig:hierarchy} shows an example account hierarchy where we label nodes with their roles (e.g., a MUA node has the M, U, and A roles). The initial node created by the genesis transaction is at the bottom. Each node is labeled with its set of roles. Each UA node represents an organization authorized to manage user accounts. The MUA nodes authorize the UA nodes and can undo any undesired action taken by the UA nodes, since they are on the path from all UA nodes to the root. This action could be taken if there is negligence on the part of a UA node in creating U nodes or if a UA node's credentials are stolen. Note that there are two MUA nodes, one on top of the other. The topmost node will be used to create and delete UA nodes, the bottom one will be used to fix the system in the event that the topmost node's credentials are stolen. This is also the reason why there are two MCLUA nodes, one on top of the other. The root node ideally is never used again after creating the MCLUA node above it. This helps prevent the root node's credentials from being stolen. In general, actions should be performed by nodes higher up in the tree that have the least privilege possible since the use of a node puts it in a more vulnerable position. The credentials of nodes not used can be secured simply by converting them to physical form and locking them in a safe (which we recommend doing with the initial node's credentials). This hierarchical node and role structure then enables the currency administrator to create a defense in depth security model. Accounts lower in the hierarchy have greater power and their credentials should be locked securely and rarely used.

A last capability not yet discussed is that accounts with M roles can issue policy that alters the cryptocurrency specification. In the event of policy conflicts between different M nodes, the nodes closer to the root are more authoritative. For M nodes the same distance from the root, those labeled with the M role in earlier blocks are more authoritative. In the event of a tie, the node labeled with the M role first within the same block wins. 

The policy deployed by the M nodes define the cryptocurrency. It is this policy that makes our approach an architecture. The policy can be set such that the cryptocurrency acts in an entirely unmanaged mode like the many popular open consensus cryptocurrencies in use today. The policy can also be set to allow the currency administrator full control as with the administrators in Multichain. More interesting to our research though is when the policy combines both open consensus and managed currency features. The policy enables each of the roles to be enabled or disabled and grants/limits the power of each role. Policy also can affect the mining community. A policy transaction can set a particular block reward or define a minimum transaction fee. Controlling these will affect the size of the mining community. For a proof-of-work based consensus mechanism such as Bitcoin, this will then indirectly control the amount of electricity used to manage the cryptocurrency (trading off power consumed against robustness of the mining pool against attack). This approach can enable an energy efficient proof-of-work consensus system where the currency administrator balances overall mining power desired vs. energy consumed. The exact capabilities available with policy are covered in section \ref{policy}.

\section{Bitcoin Specification Overview}
\label{Bitcoin Specification}
There does not exist an official Bitcoin specification. The original Bitcoin paper \cite{Nakamoto2008} contained the primary architectural details but the specification is defined by the applications that maintain it on the network. That said, there exists a Bitcoin reference client 'bitcoind' and related protocol documentation \cite{bitcoinwiki}. From this was created a useful developers reference \cite{Okupski2014}. An in depth research analysis of Bitcoin is available in \cite{Bonneau2015}.

\begin{figure*}
\centerline{\includegraphics[scale=.4]{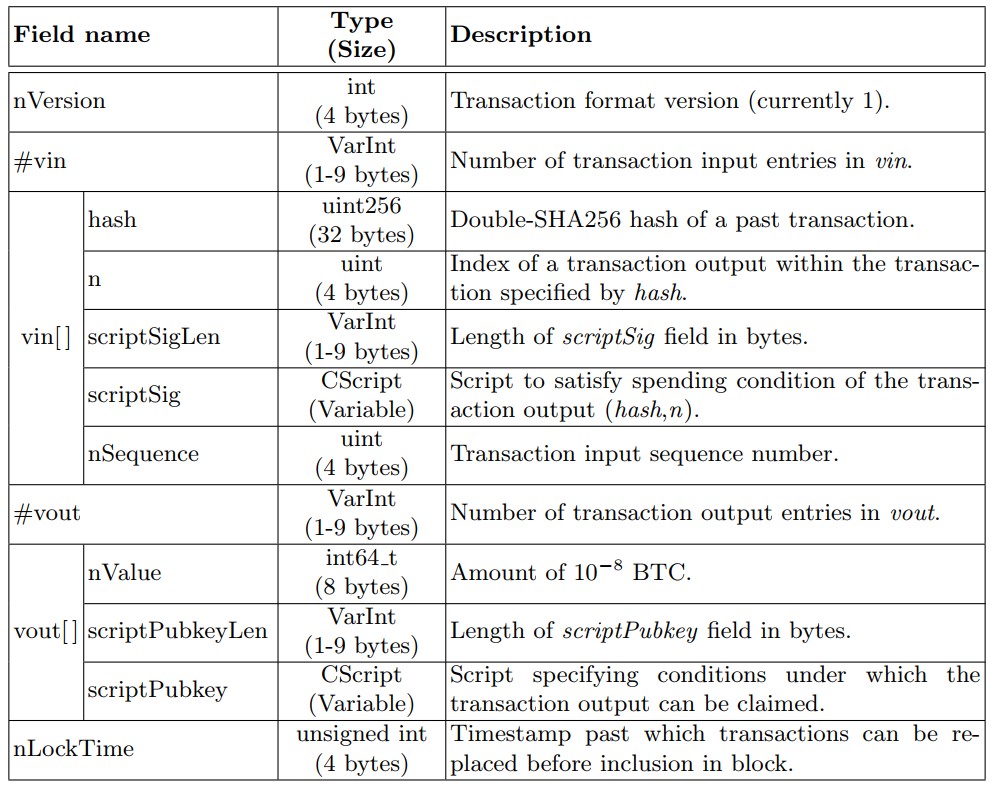}}
\caption{Bitcoin Transaction Format for Sending Bitcoin (BTC), copied from \cite{Okupski2014}.}
\label{fig:bitcoinFormat}
\end{figure*}

In this section we briefly review the features of the Bitcoin specification that will be of use for our modified specification. Figure \ref{fig:bitcoinFormat} shows the layout of a Bitcoin transaction (copied from \cite{Okupski2014}, see this for details). The vin[] sections describe the inputs to a Bitcoin transaction (the particular coins to be spent). The hash and n values specify particular coins from the output of some other Bitcoin transaction. The scriptSig is a script to provide cryptographic evidence that the owner of the coins approves of the coins being spent. It is a response script that meets the conditions of the challenge script in the transaction containing the coins that are to be spent (see the vout[] scriptPubKey field below). These conditions are usually met by proving ownership of the private key associated with the coins.

The vout[] sections describe the outputs to a Bitcoin transaction (groupings of coins along with who owns each group). Ownership is specified within each scriptPubkey which is a script defining how the coins can be spent (usually specifying a public key). To satisfy the scripPubkey challenge script and spend the coins at some future time, the owner will need to generate a scriptSig response script in some vin[] field for some transaction in which they prove ownership of the private key associated with the specified public key. This is the Pay-to-Pubkey (P2PK) Bitcoin transaction type for moving coins between accounts (see section 4.3.1 of \cite{Okupski2014} for a detailed explanation). 

\begin{figure*}
\centerline{\includegraphics[scale=.4]{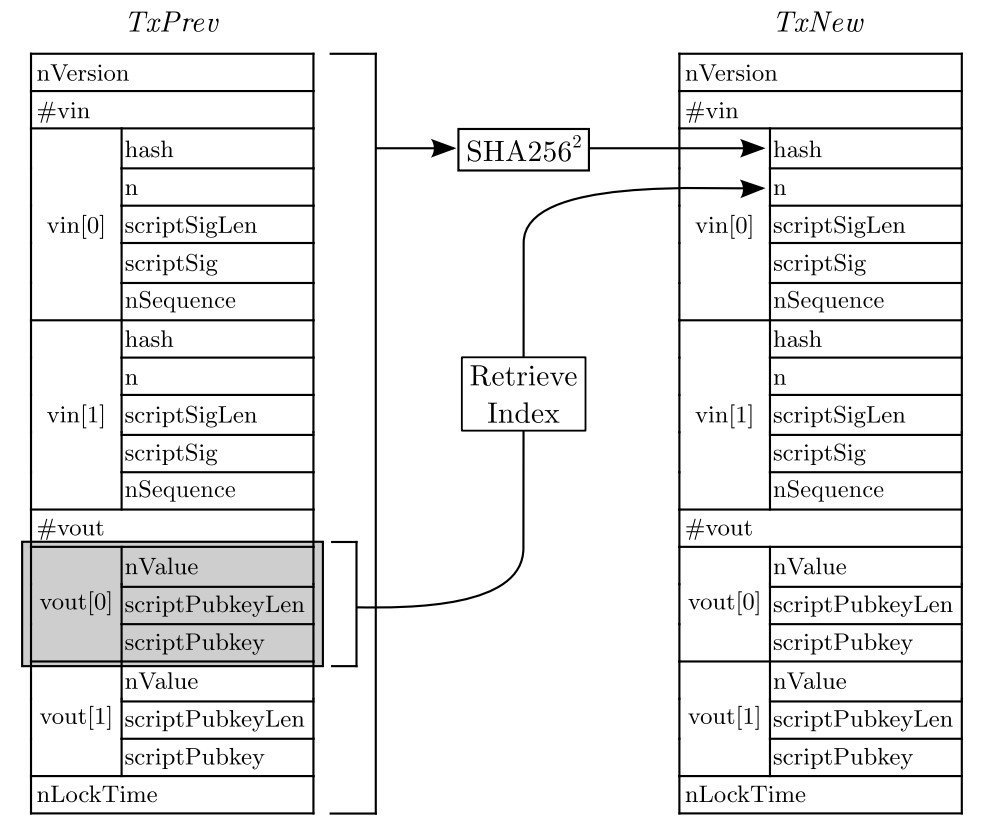}}
\caption{Bitcoin vin[] Reference to a Previous Transaction (copied from \cite{Okupski2014}).}
\label{fig:bitcoinVinRef}
\end{figure*}

Figure \ref{fig:bitcoinVinRef} shows how a vin[] field in a new transaction can reference a specific vout[] field in a previous transaction (copied from \cite{Okupski2014}, see this for details). The vin[] hash value specifies the transaction and the n value specifies the specific vout[] field. The scriptSig in the vin[] of the new transaction then satisfies the scriptPubkey from the vout[] field specified from a previous transaction so that the coins can be spent (i.e., proving that the owner of the coins wants them spent).

\section{Technical Design Using Bitcoin Specification Modifications}
\label{Technical Design}

\begin{figure*}
\centerline{\includegraphics[scale=.3]{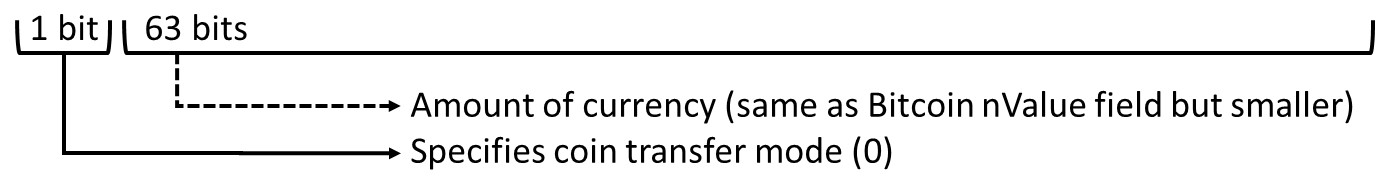}}
\caption{64 bit nValue Field Format for the Coin Transfer Mode}
\label{fig:CoinTransferFormat}
\end{figure*}

This section provides the technical specification for our managed cryptocurrency architecture described in section \ref{Architecture}. Our approach is to implement our architecture using only modest changes to the Bitcoin specification, changing the regular Bitcoin transaction format. Section \ref{Bitcoin Specification} provided the necessary background on the Bitcoin specification. Interested readers should also consult the de facto Bitcoin specifications \cite{bitcoinwiki} and \cite{Okupski2014} to better understand these changes in the context of the larger blockchain system. 

To implement our architecture's functionality, we repurpose the regular Bitcoin transaction. The format remains the same as the Bitcoin transaction shown previously in figure \ref{fig:bitcoinFormat} with a few exceptions. Our primary change is to leverage and revamp the vout[] nValue field in order to implement account roles and cryptocurrency policy. Another major change is to require in a transaction the inclusion of vin[] fields that provide the necessary roles for a transaction to be valid.

Our first modification was to change the transaction format version, nVersion, to 1944\footnote{This is the year big band leader Glenn Miller died while flying to France to encourage allied troops.}. 
Transaction format version 1 is used by the regular Bitcoin transactions and is disallowed by our architecture.

The vin[] field operates similarly as before. In Bitcoin, a vin[] field specifies a set of coins from a particular transaction already posted on the blockchain. The vin[] field then provides the evidence that the owner of those coins wants to spend them by providing a vin[] scriptSig field that satisfies the vout[] scriptPubkey field of the coins to be spent. In our design, the vin[] field works the same way for coin transfers. 

However, the vin[] field can also be used to bring roles into a transaction to authorize activities that require roles (which is most any activity in our architecture, depending upon the specific policy enacted). Functionally, it is like we are `spending' a role to use it to authorize some action given the usual use of a vin[] field (but roles can be `spent' an infinite number of times and are not transferred like coin). A vin[] field can specify a former transaction where an account was given a role. The vin[] scriptSig field then provides evidence that the owner of that account wants to use their role in this transaction (the scriptSig field must satisfy the scriptPubkey field of the transaction where the account was given the role). Thus, each vin[] field can bring a particular role from a particular account into a transaction in order to meet the role requirements for that transaction.

The vout[] field was also reinterpreted. The nValue field now specifies the mode in which its encompassing vout[] field will operate. There are three modes: coin transfer mode, role change mode, and policy change mode. Coin transfer mode moves coin between accounts similarly to a normal Bitcoin transaction. However, we restrict the transaction types that can be used in order to ensure that coins are linked to accounts. Role change mode enables accounts with the M, A, and L roles to modify the role labels of other accounts. Policy change mode enables accounts with the M role to enact and/or modify cryptocurrency policy (to essentially define the ongoing rules for the cryptocurrency). If the first bit of an nValue field is a 0, the encompassing vout[] field is in coin transfer mode. If the first two bits of an nValue field are `10', the encompassing vout[] field is in role change mode. And a nValue field beginning with `11' specifies policy change mode.

Also within the vout[] field, we restrict the scriptPubkey field to only use the Pay-to-Pubkey (P2PK) transaction type. P2PK associates coins with a specific public key (an account in our architecture). If set up to do so, this enables cryptocurrencies implemented from our architecture to link accounts to account owners. This linkage can take place when an account with the A role grants the U role to another account (thereby authorizing it for coin transfers). In this case, the authorizing entity checks the user's identity using out-of-band traditional methods (e.g., passports, drivers licenses, and identity cards).

\subsection{Coin Transfer Mode}

If an nValue field has its first bit set to 0, the encompassing vout[] field is in coin transfer mode and is used to move coin between accounts. Since the first bit was used to specify this, the remaining 63 bits specify the amount of coin to be transferred (in Bitcoin all 64 bits are used). Figure \ref{fig:CoinTransferFormat} shows the changes to the nValue field for the transfer of coin (those nValue fields beginning with 0). Note that for all figures showing the revised nValue format (including this one), solid lines originate from bits that define the action to be taken while dotted lines originate from parameter values.

Anytime a transaction has one or more vout[] fields in coin transfer mode, the original accounts owning the coins and the destination accounts for the coins must all have the U role. This is accomplished by including in the transaction vin[] fields that bring in the U roles for the accounts either sending or receiving coin. 

Lastly, coinbase transactions (the first transaction of each block where the miner sends itself the reward coins) are handled the same as with Bitcoin. However, the vout[] nValue field will start with a 0 bit, putting it in coin transfer mode. Also, the miner must include a vin[] field after the normal coinbase transaction vin[] field in which the miner provides the U role for the account to which the coins are destined. 

\subsection{Role Change Mode}

\begin{figure*}
\centerline{\includegraphics[scale=.3]{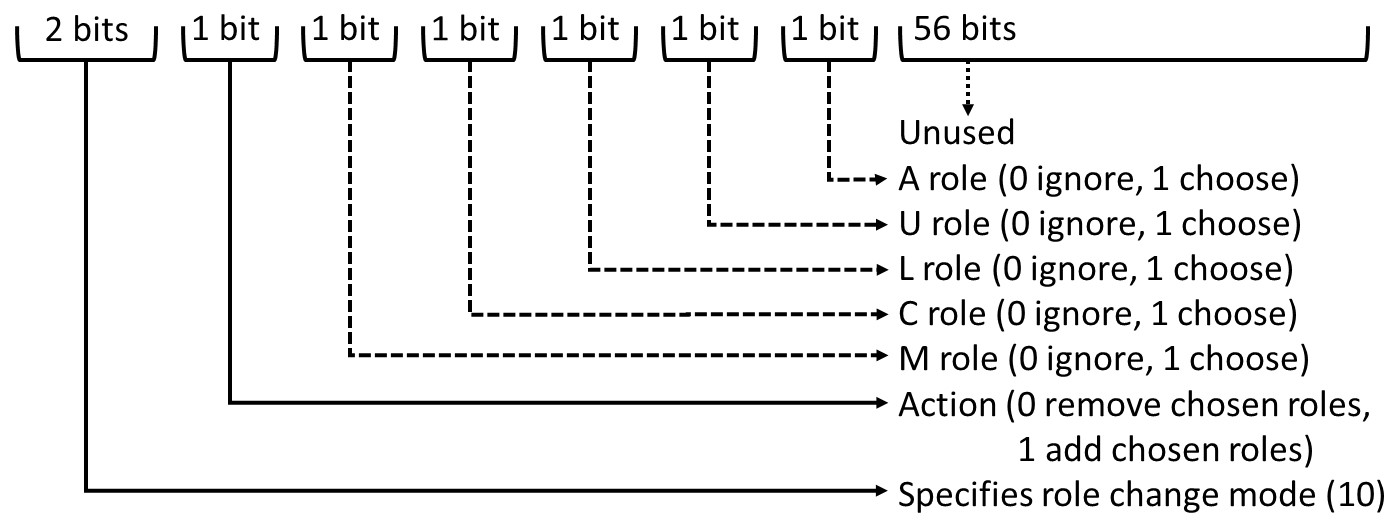}}
\caption{64 bit nValue Field Format for the Role Change Mode}
\label{fig:RoleChangeFormat}
\end{figure*}

If an nValue field has its first two bits set to `10', then the encompassing vout[] field is used to change the roles for a set of accounts. The third bit represents whether or not the vout[] field is removing or adding roles. 0 indicates that roles are being removed and a 1 represents that they are being added. The subsequent bits are flags referring to the different roles. Bits 4, 5, 6, 7, and 8 map to roles M, C, L, U, and A respectively. The remaining 56 bits are undefined. This may be wasteful of space but role change transactions will be relatively rare and we are trying to change the Bitcoin specification as little as possible. Figure \ref{fig:RoleChangeFormat} shows these changes to the nValue field.

The vout[] scriptPubkeyLen and scriptPubkey fields specify the public key for the account that has these roles. The roles granted by the transaction can then be used in future transactions by the future transaction providing a vin[] scriptSig field that satisfies the vout[] field of the transaction granting the roles. Essentially, an owner of an account uses their private key in some future transaction to prove ownership of a public key documented in a past transaction where the roles were granted. Note that cryptocurrency participants, specifically the miners, will have to make sure that the roles being accessed by a transaction haven't been previously removed from the relevant accounts (roles can be removed by accounts with the M, L, or A roles). This check is similar to miners in Bitcoin checking to make sure that particular coins haven't already been spent.

Every transaction requires one or more roles in order to be valid. Each role has different rules that must be satisfied for the applicable transaction to be valid:

\subsubsection{M Role Processing}
\label{subsub M role}
Any addition or removal of roles requires the M role to be provided in one or more of the vin[] datastructures (with two exceptions, see the A and L roles). Each role change vout[] datastructure must be `covered' by a vin[] scriptSig field where the address specified is located between the root and the node affected in the node hierarchy. Also, the `covering' address (referenced by the vin[] scriptSig field) must have the role that is to be added or removed in the `covered' vout[] datastructure.

\subsubsection{C Role Processing}
The inclusion of a vin[] datastructure that has a scriptSig field that satisfies an account having the C role means that the transaction may create coins. There is no need then for other vin[] datastructures. The vout[] datastructures provide coins to the designated addresses.

\subsubsection{L Role Processing}
The inclusion of a vin[] datastructure that has a scriptSig field that satisfies an account having the L role means that the other vin[] fields do NOT need the scriptSigLen or scriptSig fields (for bringing coin into the transaction). Coins may be transferred without the permission of the owners with the inclusion of the L role in the transaction. Also, having the inclusion of the L role enables vout[] datastructures that remove the U role from other accounts. Also, the U role may be added back to accounts for which it was previously revoked. However, these abilities only apply to nodes in the hierarchy that are at a greater distance from the root than the vin[] specified node with the L role (this is to enable the currency administrator to limit this power by creating L role accounts at differing distances from the root).  

\subsubsection{U Role Processing} 
Any movement of funds requires the U role for the original owner of the coins (specified in the vin[] fields). The recipients of any coins (specified in the vout[] field) must also have the U role.

\subsubsection{A Role Processing}
The inclusion of a vin[] datastructure that has a scriptSig field that satisfies an account having the A role means that the vout[] fields may add role U to accounts. Doing so adds them as descendants in the hierarchical account tree. Accounts with the A role may likewise remove the U role from any descendant. If an A node removes one of its descendants U roles, another A node may add the U role to that node. In this case, the affected node becomes a descendant of the A node adding the U role. Note that if a node with the L role removes the U role from a node, it is put on a special list of frozen nodes and only another node with the L role may remove the affected node from the list.

\subsection{Policy Change Mode}
\label{policy}

\begin{figure*}
\centerline{\includegraphics[scale=.3]{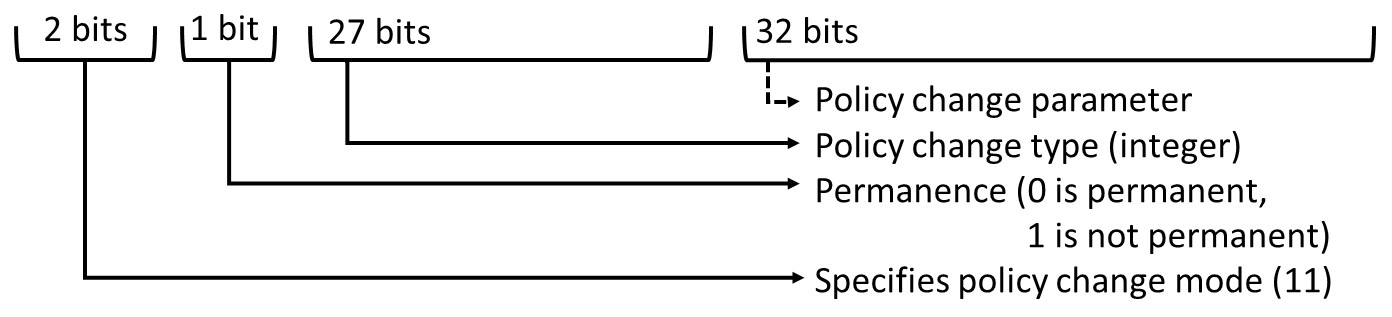}}
\caption{64 bit nValue Field Format for the Policy Change Mode}
\label{fig:PolicyChangeFormat}
\end{figure*}

If an nValue field has its first two bits set to `11', then the encompassing vout[] field is in policy change mode, used to create or modify cryptocurrency policy. Note that a vout[] field in policy change mode is only allowed in a transaction if at least one of the vin[] fields provides the M role (since only currency managers can modify policy). 

The third bit of the nValue field defines the permanence of the policy (0 is not permanent and 1 is permanent). If an account issues permanent policy, it may not change it in the future. However, M accounts with greater priority, as described in section \ref{Architecture}, can still trump the issued policy. If the initial root node issues permanent policy, it cannot be changed for the life of the cryptocurrency. This enables the issuance of a static instance of our cryptcurrency architecture. Some features may be made permanent while others are left open for change. It may not be immediately clear why an issuer of a currency would make anything permanent, because it reduces their flexibility. However, by making certain features permanent it provides guarantees to the users. The currency administrator is then constrained to operate within the published rules of the cryptocurrency even though they still manage it. This idea of permanence is important in order to limit the currency administrator from having absolute rule (which is the case in many of the private blockchain managed cryptocurrencies, such as with Multichain \cite{Greenspan2015}).

After the first three bits of an nValue field are set (to 110 for not permanent or 111 for permanent), the remaining 61 bits specify the policy setting to be made. There is just one policy change made per nValue field, and just one nValue field per vout[] datastructure. However, a single transaction may have many vout[] datastructures.

The next 27 bits specify an integer representing the policy change type while the last 32 bits are used to hold the policy change parameter. The structure of the nValue field in the policy change mode is shown in figure \ref{fig:PolicyChangeFormat}.

\begin{table*}[htbp]
  \caption{Cryptocurrency Policy Settings}
  \label{tab:policy}
  \begin{center}
  \begin{tabular}{ccl}

Policy Change Type & Description & Parameter \\
\hline                                                
0  & Enable or disable the M role globally                          & 0 or 1    \\
1  & Enable or disable the C role globally                          & 0 or 1    \\
2  & Enable or disable the L role globally                          & 0 or 1    \\
3  & Enable or disable the U role globally                          & 0 or 1    \\
4  & Enable or disable the A role globally                          & 0 or 1    \\
5  & Enable or disable the L roles from moving coins                & 0 or 1    \\
6  & C role coin creation limit per block (0 means no limit)        & Integer   \\
7  & Set block reward mode (0 means manual, 1 means self-adjusting) & 0 or 1    \\
8  & For manual mode, set block reward                              & Integer   \\
9  & For manual mode, set minimum block reward                      & Integer   \\
10 & For self-adjusting, set geometric decay rate					& Float between 0 and 1 \\
11 & For self-adjusting, set maximum decay rate                     & Float between 0 and 1 \\
12 & Set transaction fee minimum (0 means no minimum)               & Integer   \\
13 & Periodicity of management transaction inclusion in blocks      & Integer   \\
14 & Minimum number of management transactions per period           & Integer   \\
15 & No operation (used to prove the currency administrator is active)                 & 0         \\

  \end{tabular}
  \end{center}
\end{table*}

For the policy change mode, there are currently 14 policy change types with associated parameters, shown in table \ref{tab:policy}. For the binary parameters, 0 means disable and 1 means enable. Binary parameters default to 1 (these policies are enabled by default when the cryptocurrency is initiated).  

Policy change types 0 to 5 enable or disable the various roles in available in the architecture (discussed in section \ref{Architecture}). Type 5 enables or disables the L role from moving coins (disabling would limit the L role to freezing accounts). Type 6 sets a limit for how much coin the set of C roles may create within any particular block. Type 7 sets the block reward mode (0 is the automated approach used by the base cryptocurrency system, Bitcoin in our case, while 1 enables a mode where a currency manager explicitly sets rewards). Type 8 and 9 are for the manual mode and enable setting the block reward and setting a minimum block reward. The purpose of the type 9 is to allow a currency manager to permanently set a minimum while still having the flexibility to adjust the current reward with type 8. Types 10 and 11 are for the self-adjusting mode and enable setting the decay rate for block rewards as well as setting a maximum decay rate. Again, the latter is intended to be used in a mode where it is set permanently. Type 12 sets a transaction fee minimum. 

Types 13-15 are important for setting security policy (discussed in detail in section \ref{Security}). Type 13 sets how often management transactions must appear in a consecutive sequence of blocks (0 disables this feature). For example, a setting of 5 indicates that a certain number of management transactions must appear within every subsequent grouping of 5 blocks. Type 14 specifies the minimum on how many management transactions must appear in that grouping of blocks. A management transaction is one that requires the M role to be present in one of the vin[] fields (see section \ref{subsub M role}). If the currency administrator doesn't have enough management transactions that they wish to put on the blockchain to meet the minimum, then they may issue one or more no operation (no-op) policy change mode transactions of type 15 using one of their M nodes. These do nothing but meet the requirement. A last nuance of this mechanism is that at least one of the management transactions must be a policy change mode transaction. This is to ensure that the currency administrator can always change policy (as the miners might just include non-policy management transactions to meet the minimum requirement). 

\section{Security Models}
\label{Security}
A key aspect of our architecture is to ensure that a balance of power is maintained. Users of the system, including currency managers, should be able to issue any valid transaction onto the blockchain (pursuant to the current policy settings). Miners should be able to enforce policy restrictions and provide transparency for all transactions added to the blockchain.

There are two security models that can be used to enforce this balance of power. Each model slightly favors one party, currency managers or miners, although both achieve a reasonable balance (dependent upon the use case). 

\subsection{Independent Mining Model}
In the independent mining model, the currency administrator permanently disables the requirement to include management transactions periodically (thus the blockchain is not dependent on receiving management transactions). This can be done by having the initial node permanently set the policy change type 13 to 0. In this mode the currency administrator cannot take over maintenance of the blockchain (since mining is unrestricted as with Bitcoin). However, if at least 51 \% of the miners collude to `revolt' against the currency managers, they can prevent future management transactions from entering the blockchain (as well as issuing the well known set of 51 \% attacks present with most blockchains \cite{yli2016current}). The way this attack works is that the miners controlling 51 \% of the computational power simply work on a chain with only their own blocks, excluding the blocks produced by others. Over time, their chain will be longer since they own the majority of the computational power and the other miners will follow their chain (fruitlessly trying to append blocks in a competition they will never win)

\subsection{Dependent Mining Model}
Even though the 51 \% attack possibility exists in Bitcoin and most other cryptocurrencies, the risk may be too great for some issuers of cryptocurrency; in such a case, the currency administrator can use our dependent mining model. In this case the blockchain is dependent on receiving management transactions. With this approach, the currency administrator using an M node sets policy change types 13 and 14. This forces the miners to include a certain number of management transaction per a certain number of blocks. We advise setting this liberally (type 13 large and type 14 small) since the expectation is that 51 \% of the miners will not revolt. If a revolt occurs and miner only include the minimum necessary, then these policy values can be changed to force the miners to allow for more management transactions.

If the miners completely revolt and violate policy, the `compliant' miners will reject their blocks. This would fork the blockchain into a compliant chain and a non-compliant chain. This is the same thing that would happen with any cryptocurrency if a group of miners begin producing blocks that do not satisfy the specification requirements.

An important aspect of this second model is that it gives more power to the currency administrator than the first model. This can be seen as a positive feature or a weakness depending upon the use case and perspective. With the second model, the currency managers accounts can refuse to submit management transactions, which will eventually cause block creation to halt (issuing management transactions would immediately restart production). This may not be considered a significant threat as the currency administrator initiated the blockchain and inherently will want it to continue operating (this argument is somewhat analogous to the one explaining why Bitcoin in practice is resistant to a 51 \% attack even though theoretically it is vulnerable \cite{yli2016current}: the miners have a huge stake in the system and won't want it to fail). This could even be considered a feature as owners of a blockchain could eventually deprecate it and move the data to a new blockchain with enhanced technical capabilities. Note that using such an option would be extremely visible and necessarily be rare as it would require all of the users' cryptocurrency software to be updated and reconfigured.

\subsection{Node Software Security}
We should note that in all cryptocurrency systems, the authors of the software used by the participating nodes (especially the mining nodes) have significant power. Our architecture is no exception. However, here there is also a balance of power. The currency administrator will likely be a maintainer of the software used by nodes to maintain the blockchain. Hypothetically, they could use this to violate established permanent policy and/or take control of the blockchain from the miners through the creation and publication of `malicious' software. However, this can only occur if the majority of miners adopt the malicious software. Even if this did happen (e.g., through miners blindly adopting an update), the miners could simply roll back to a previous non-malicious version to restore the proper function of the architecture.

If miners author the node software, they publish `malicious' software, and the majority of miners adopt it, the miners could revolt against the currency administrator. However, this is identical to a 51 \% attack as described above. The result would be a forking of the blockchain, creating compliant and non-compliant chains. The compliant chain would continue to implement our architecture with a reduce set of compliant miners.

%\section{Analysis}
%\label{Analysis}
\section{Conclusion}
\label{Conclusion}

We provide a novel cryptocurrency architecture which is a hybrid approach where a managed cryptocurrency is maintained through distributed open consensus based methods. Key to this architecture is the idea of a genesis transaction upon which all other transactions are based and which enables the establishment of a hierarchy of accounts with differing roles. It is these roles that enabled us to introduce features from fiat currencies into a cryptocurrency: law enforcement, central banking, and account management. Another novel feature is that the architecture allows the cryptocurrency policy to be maintained dynamically by the currency administrator, but certain policy settings can be made permanent in order to facilitate confidence in the stability of the system. This is especially important for the relationship between the currency administrator and an independent community of miners. The currency administrator can control block rewards, which indirectly enables the currency administrator to adjust the power consumption of blockchain maintenance. However, the currency administrator can enact permanent policy to guarantee the miners a certain level of reward. This is important not only to the miners but it prevents the currency administrator from lowering the block reward to nothing and then taking over the mining (and thus completely controlling the blockchain as with many permissioned blockchain systems). Our policy system thus enables a cryptocurrency to be set up that has a balance of power where the currency administrator can perform management functions but where a group of independent miners enforce policy and provide transparency through recording all administrative activity on the blockchain. However, the possibility still exists that the currency administrator or miners could violate policy and attempt to take control of the system. To mitigate this, we provide two security policies that can enforce the balance of power (each with a small bias one direction or the other). Lastly, we showed that our architecture can be implemented through modest changes to the Bitcoin specification. We note though that our approach is not tied to Bitcoin and can be implement on differing cryptocurrency platforms.

%\section{Conclusion}
%\label{Conclusion}

%We have shown how to combine advantages of open consensus based cryptocurrencies with features from managed currencies in an architecture that can be used to issue different types of currencies. The architecture eliminates the need for users to completely trust the currency administrator, but at the same time we have enabled the currency administrator to manage the cryptocurrency. We leveraged the idea of a genesis transaction to label accounts with roles, which are then critical for the management of the currency. We then set up the miners to enforce the rules of the cryptocurrency as well as the policy issued by the currency administrator (most critically the issued `permanent' policies that define the cryptocurrency instance). To make this work, we had to evaluate how to maintain a balance of power between a public group of miners and the currency administrator. Among many other capabilities, we have enabled the currency administrator to control the overall power consumption of the system. We identified the strengths and weaknesses of this hybrid control approach and demonstrated how it could be implemented through modest modifications of the popular Bitcoin protocol

\bibliographystyle{IEEEtran}
\bibliography{IEEEabrv,MCpaper}

\end{document}